\documentclass[12pt,oneside,british]{amsart}
\usepackage{mathptmx}

\usepackage[T1]{fontenc}
\usepackage[utf8]{luainputenc}
\usepackage{geometry}
\usepackage{amsthm}
\usepackage{amssymb}
\usepackage{setspace}
\usepackage[authoryear]{natbib}
\makeatletter
\numberwithin{equation}{section}
\numberwithin{figure}{section}
  \theoremstyle{definition}
  \newtheorem{defn}{\protect\definitionname}
\theoremstyle{plain}
\newtheorem{thm}{\protect\theoremname}
  \theoremstyle{plain}
  \newtheorem{cor}{\protect\corollaryname}

\makeatother

\usepackage{babel}
  \providecommand{\definitionname}{Definition}
\providecommand{\corollaryname}{Corollary}
\providecommand{\theoremname}{Theorem}

\begin{document}

\title{Economics cannot isolate itself from political theory: A mathematical
demonstration}

\author{Brendan Markey-Towler}
\begin{abstract}
The purpose of this paper is to provide a confession of sorts from
an economist to political science and philosophy. A confession of
the weaknesses of the political position of the economist. It is intended
as a guide for political scientists and philosophers to the ostensible
policy criteria of economics, and an illustration of an argument that
demonstrates logico-mathematically, therefore incontrovertibly, that
any policy statement by an economist contains, or is, a political
statement. It develops an inescapable compulsion that the absolute
primacy and priority of political theory and philosophy in the development
of policy criteria must be recognised. Economic policy cannot be divorced
from politics as a matter of mathematical fact, and rather, as Amartya
Sen has done, it ought embrace political theory and philosophy.
\end{abstract}

\address{School of Economics, University of Queensland, Australia}

\email{brendan.markeytowler@uqconnect.edu.au}


\maketitle

\section{The place and importance of Pareto optimality in economics}

Economics, having pretensions to being a ``science'', makes distinctions
between ``positive'' statements about how the economy functions
and ``normative'' statements about how it should function. It is
a core attribute, inherited largely from its intellectual heritage
in British empiricism and Viennese logical positivism \citep{McCloskey1983}
that normative statements are to be avoided where possible, and ought
contain little by way of political presupposition as possible where
they cannot. Political ideology is the realm of the politician and
the demagogue.

To that end, the most basic policy decision criterion of Pareto optimality
is offered. This criterion is weaker than the extremely strong Hicks-Kaldor
criterion. The Hicks-Kaldor criterion presupposes a consequentialist,
specifically utilitarian philosophy and states any policy should be
adopted which yields net positive utility for society, any compensation
for lost utility on the part of one to be arranged by the polity,
not the economist, out of the gains to the other. Against such a criterion
which clearly washes its hands of the actions of powerful entities
in the polity, alongside the standard problems of utilitarianism \citep{Sen1973,Fumagalli2013},
Pareto optimality augurs only that any policy be adopted which leads
to an at least indifferent state for all, and an improved state for
at least one.

Aside from the presupposition (again) of a consequentialist philosophy,
this appears quite a ``weak'' dictat, requiring not much by way
of political presuppositions. It says nothing about who is to ``lose''
from policy, and only concerns ``gain''. And yet such a weak dictat
allows the economist to claim that by removing impediments to the
perfect market (``market imperfections''), allowing laissez-faire
competition in markets free of extra-judicial governmental intervention,
yields an ``optimal'' or ``efficient'' outcome for society. Because
the first and second ``welfare theorems'' tell us (roughly) that
within the strictures of the psychological model of the neoclassical
rational agent, market ``equilibria'' are Pareto optimal \citep{Mas-Collel1995}:
no individual can be made ``better off'' without making some other
individual ``worse off''.

If we restrict what constitutes a political statement to one which
makes statements about when policies should be implemented which will
lead to the \emph{dis}improvement of some individual's situation.
Then economics appears to have made only a value judgement about who
ought benefit from a policy (whosoever accrues such benefits). Not
(by this restricted definition) a political statement. And yet it
has still demonstrated markets free from government intervention,
and free from imperfections are ``optimal'' and ``efficient''.

But is the concept of Pareto optimality robust? Does it have any value
as a criterion in the ``real'', empirical world? Does it offer us
a criterion for policy which does not make political statements as
defined here, and allow for economics to be divorced from political
theory? Indeed assert its priority and primacy therein?

The purpose of this work is to demonstrate logico-mathematically,
therefore incontroveritbly that the answer is No. The mathematics
of Pareto optimality itself provide an inescapable argument which
compels us to recognise that even in the restricted form here, economics
cannot but make political statements. Economics must recognise the
absolute primacy and priority of political theory and philosophy in
the development and implementation of policy.

The author is not a political scientist or philosopher, and offers
no pretensions to being esteemed as such or even to being versed in
the literature to such an extent as one would be. The author is an
economist, neoclassically trained, and is offering a confession of
the weakness of the political position held by the economics profession.
What is offered is a guide for political scientists and philosophers
to the exactitudes of the ostensible policy criterion of the economist,
and an illustration, a mathematically rigorous and therefore incontrovertible
illustration that this criterion offers no guide in empirical situations.
That even political statements of a restricted nature must \emph{always}
be made by economists designing and implementing policy, and thus
the concomitant compulsion to embrace political theory and philosophy
as prior to any analysis of economic policy.

This confession ought not be read as purely negative. It is as much
an affirmation of Professor Sen's long-time collaboration with Professor
Nussbaum, embracing political theory and philosophy as a means for
developing a welfare economics with the intellectual richness of its
foundation in the same, as it is a critique of economics. It ought
be read as encouragement for \emph{both} economists and political
scientists and philosophers to continue the necessary development
of Sen and Nussbaum's endeavour.

The argument proceeds as follows. In the first succeeding section
we consider some matters of definition as regards the weakest possible
conception of Pareto optimality. In the second section we relate this
weakest form to that actually employed in neoclassical economics for
policy analysis (specifically, the defence of \emph{laissez-faire}
free markets corrected for ``imperfections''). And we show how concerns
begin to emerge as this criterion seems to deem as ``efficient''
or ``optimal'' extreme situations. Before we demonstrate, in the
final substantive section, that in \emph{all }non-extreme empirical
situations, \emph{all} states of the world are Pareto optimal, so
that \emph{any} policy analysis in empirical reality must necessarily
make political statements, \emph{even} in the restricted sense of
those adopted here.

Finally, before we begin proper, we might do well to ask (and we will
return again to the answer below); so what? Why do we care? Why should
we care?

We should care that economics cannot usurp the primacy of political
theory and philosophy in policy analysis because the delineation of
economic science from political economy, the delineation of that fuzzy
boundary between fact and value \citep{Strauss1953}, the seeking
of ``objectivity'' is essential to a healthy political sphere \citep{Sen1993}.
Something we may all agree is desperately needed given the current
political situation in the democracies of the world. Undergraduate
economists are still taught the concept of Pareto optimality as the
basis for economic policy, professional economists still utilise it
in research, it still forms the basis for the ``proof'' that laissez-faire
markets (corrected for ``imperfections'') are ``efficient'' or
``optimal''. But there are no ``ought'' statements to be derived
by the economist \emph{qua} economist devoid of political content,
political presuppositions. To continue to pretend otherwise, by refusing
to embrace political theory and philosophy and acknowledge the primacy
and priority of political concerns in policy implementation lends
to the pronouncements of the economist a false scientistic authority
detrimental, even dangerous, for the process of public reasoning.

\section{The concept of Pareto efficiency in its weakest form}

The following conceptualisation of Pareto efficiency is a weaker form
of the concept used in economic theory \citep{Mas-Collel1995,Sen1970},
because individual preferences are defined with respect to arbitrary
sets of information contained within the state of society. We assume
only that individuals care about Something. Not one particular Thing
such as their acquisition of commodities.

We first take the polity:
\begin{defn}
$N$ is the set of all individuals in society.
\end{defn}
And that which their politics concerns - the state of society.
\begin{defn}
$S$ is the set of all possible information contained within society,
so that a set $s\in2^{S}$ ($2^{S}$ being the set of all possible
subsets of $S$) contains all extant information about a particular
iteration of society and will be called the \emph{state of society}.
$S$ is an arbitrary topological space.
\end{defn}
And the means by which individuals make judgements about that which
their politics concerns. Their preferences over the information contained
within the state of society.
\begin{defn}
Each individual $i\in N$ has a complete and transitive preference
relation $\succeq_{i}$ defined over a set of preference-information
$S_{i}\subset S$ such that $s_{i}\succeq s_{i}'$ can be read ``individual
$i$ prefers preference information $s_{i}$ at least as much as preference-information
$s_{i}'$''.
\end{defn}
Any particular set of preference-information $s_{i}\subset S_{i}$
can be thought of as the state of society as \emph{viewed }by individual
$i$. The set of preference-information for individual $i$ is a subset
of the information contained within a particular iteration of society,
so $s_{i}\subset s\subset S$.

A particular state of society $s$ is a Pareto efficient if there
is no other state of society $s'$ for which one individual strictly
prefers their preference-information $s_{i}'\subset s'$ to that particular
state $s_{i}\subset s$, and the preference-information $s_{j}'\subset s'$
in the other state $s'$ is at least as preferred by every other individual
$j\ne i$.
\begin{defn}
A state $s\in S$ is said to be \emph{Pareto efficient }if and only
if $\nexists s'\in2^{S}\,\&\,i\in N:s_{i}'\succ s_{i}\,\&\,s_{j}'\succeq s_{j}\,\forall\,j\ne i\in N$.
\end{defn}
To put it crudely, a particular state of society is Pareto efficient
if no individual can be made ``better off'' without making another
individual ``worse off''. A dynamic concept which mirrors this (and
we will see provides an alternative definition of Pareto optimality)
is the concept of a Pareto improvement - whereby a change in the state
of society leaves everyone at least indifferent, and at least one
individual in a preferable situation.
\begin{defn}
A movement between two states of society, $s\rightarrow s'$ is called
a \emph{Pareto improvement} if and only if $\exists i\in N:s_{i}'\succ s_{i}\,\&\,s_{j}'\succeq s_{j}\,\forall\,j\ne i\in N$.
\end{defn}
Note that this does \emph{not} imply that $s'$ is a Pareto efficient
state, because the same could potentially be said of a movement $s'\rightarrow s''$.
The state $s'$ is only a Pareto efficient state if we cannot find
yet another state for which the movement to that state is a Pareto
improvement. The following Theorem, quite well known, demonstrates
this distinction and gives an alternative definition of Pareto efficiency.
\begin{thm}
\label{prop:Pareto_efficiency=00003DNo_Pareto_Improvements}A state
$s\in2^{S}$ is \emph{Pareto efficient} if and only if there is no
other state $s'$ for which the movement $s\rightarrow s'$ is a \emph{Pareto
improvement}.
\end{thm}
If one adheres to a consequentialist political doctrine (such as classical
utilitarianism) rather than a deontological doctrine (such as liberalism)
in which action is guided by some categorical imperative other than
consequentialism, the guide offered by Pareto improvement is the least
controversial, and least politically committal criterion to decision-making
one can find. Indeed if we restrict political statements to those
which concern the assignation of \emph{losses}, it is a-political.
It makes a value judgement only about who ought \emph{gain} (whosoever
stands to).

Unless one holds a strict deontological doctrine in the style, say,
of \citet{Nozick1974} (in which the maintenance of individual freedom
is the categorical imperative), or \citet{Rawls1971} (in which again
individual freedom is the primary categorical imperative and the betterment
of the ``poorest'' the second categorical imperative), it is more
difficult to argue against implementing some decision which will cause
a change of society which all individuals in society will be at worst
indifferent to. Than arguing for some decision rule which will induce
a change of society which some individual will find \emph{less} preferable.
To the rationalisitic economist it seems almost petty, certainly irrational
to argue against this criterion, like those individuals who demand
``fairness'' in the famous ``dictator'' experiment rather than
accept someone else becoming ``better off'', and themselves no ``worse
off''.

\section{The concept of Pareto Efficiency in welfare economics}

Now we will turn to the concept of Pareto efficiency employed in its
far stronger form in welfare economics. The economic system is a social
system in which commodities are exchanged. Sets of these commodities
can be represented by vectors $x$ within a metric space $X$ contained
within the non-negative orthant of an Euclidean space $\mathbb{R}_{+}^{N_{x}}$
of dimensionality $N_{x}$ equal to the number of such commodities\footnote{For the unindoctrinated, imagine $N_{x}$-many rulers laid out next
to each other, $\mathbb{R}_{+}^{N_{x}}$ are these rulers, $x$ a
set of points marked off on each of them.}.
\begin{defn}
An allocation $\left\{ x_{i}\right\} _{i\in N}\subset X\subset\mathbb{R}_{+}^{N_{x}}$
of commodities in society is a set of vectors $x_{i}$ representing
the commodities allocated within the economic system to each individual
$i\in N$.
\end{defn}
In questions of welfare economics at least in all practical policy
matters, the state of society is equated with this allocation, that
is, $s=\left\{ x_{i}\right\} _{i\in N}$, and the set of all possible
information concerning the economic state of society is $S=X$. It
is typically taken to be the case that the individual's preference-information
is simply their allocation $x_{i}$, $s_{i}=x_{i}$. The concept of
Pareto efficiency is thus narrowed from that above to what we may
call ``neoclassical Pareto efficiency'' for the school of economic
thought in which originates, and to distinguish it from the weaker
criterion.
\begin{defn}
An allocation $\left\{ x_{i}\right\} _{i\in N}$ is said to be \emph{neoclassical
Pareto efficient} if and only if $\nexists\left\{ x_{i}\right\} _{i\in N}\subset X\,\&\,i\in N:x_{i}'\succ x_{i}\,\&\,x_{j}'\succeq x_{j}\,\forall\,j\ne i\in N$.
\end{defn}
This is consistent with the definition given by standard economics
texts, being a preference-axiomatic form statement of Pareto optimality
in Definition 10.B.2 by \citet[p.313]{Mas-Collel1995}. The concept
of Pareto improvement can be narrowed to ``neoclassical Pareto improvement''
in the same manner.
\begin{defn}
A movement between two allocations, $\left\{ x_{i}\right\} _{i\in N}\rightarrow\left\{ x_{i}'\right\} _{i\in N}$
is called a \emph{neoclassical Pareto improvement} if and only if
$\exists i\in N:x_{i}'\succ x_{i}\,\&\,x_{j}'\succeq x_{j}\,\forall\,j\ne i\in N$.
\end{defn}
For technical reasons it is almost always in practice assumed for
simplicity that individual preference relations are monotonically
increasing across the space of commodities.
\begin{defn}
If individual preferences are monotonically increasing then $x_{i}'\succeq_{i}x_{i}\iff x_{i}'\ge x_{i}$,
and $x_{i}'\succ x_{i}\iff x_{i}>x_{i}'$\footnote{Since $x_{i}$ is a vector, $x_{i}>x_{i}'$ acquires the special meaning
that each element of $x_{i}$ is at least as large as $x_{i}'$ and
at least one element is strictly greater.}.
\end{defn}
This is problematic, because a normative economics guided by the principle
of implementing a decision if it yields a neoclassical Pareto improvement
where individuals have such preference relations above leads to the
following situation.
\begin{thm}
\label{Prop:The-extreme_case}Suppose that individual's preference-information
is their own allocation of commodities, and that their preferences
are monotonically increasing. Take one individual $j\in N$ and an
initial allocation $\left\{ x_{i}\right\} _{i\in N}$.

- A series of movements between allocations $\left\{ \left\{ x_{i}\right\} _{i\in N}^{t}\rightarrow\left\{ x_{i}'\right\} _{i\in N}^{t}\right\} _{t=1}^{T}$
such that $x_{i\ne j}=x_{i\ne j}'\,\forall\,t$ and $x_{j}'>x_{j}\,\forall\,t$
and therefore that $x_{j}-x_{i}\rightarrow\infty\,\forall\,i\ne j\in N$,
are neoclassical Pareto improvements. 

- Furthermore, if these movements are made possible only by the discovery
of new commodities, each individual state in the movement is neoclassical
Pareto efficient prior to the next discovery if the first allocation
was neoclassical Pareto efficient.
\end{thm}
Admittedly perhaps not to the economic theorist, but to most this
seems a rather dubious outcome. It means that if we are guided by
neoclassical Pareto efficiency it is acceptable, indeed desirable,
that one individual within society be made increasingly ``richer''
without end and without increasing the wealth of others. Provided
only the wealth of others does not decrease. The same result would
hold if instead of an individual, we made a whole group, or indeed
the whole of society ``better off'', without making anyone else
``worse off''.

Even the most devoted disciple of Ayn Rand would find this situation
dubious, for there is no requirement that the individual in question
be in some sense ``deserving'' of their riches. But it is perfectly
logically consistent with Pareto optimality if individual preferences
concern only to their allocation and are monotonically increasing.
So what is it that is strange here? What generates this odd condonation?
It is the narrowing of that which the polity care about to each individual
allocation, alone, independent of others. The fact that neoclassical
Pareto improvements are distribution-invariant because the polity
is supposed to care only about their own individual allocation $x_{i}\in\left\{ x_{i}\right\} _{i\in N}^{t}$
alone rather than broader states of society $s_{i}\subset s$ as they
see it.

To avoid such awkward results, the economist may move from the preference-axiomatic
concept of Pareto efficiency to embrace utilitarianism. The policy
criterion (actually not immediately representative of Bentham's surprisingly
subtle statement) being the maximisation of some combination $W\left(x\right)=W\left(\left\{ u_{i}\left(x_{i}\right)\right\} _{i\in N}\right)$
of individual utilities $u_{i}\left(x_{i}\right)$ over allocations.
The ``social psychic wellbeing'' metric known as the Social Welfare
Function.

In theory, the maximisation of $W\left(x\right)$ would, given the
``right'' assumptions on the combination method $W\left(\cdot\right)$
(sum, multiplication, maximin etc.) and utilities (concavity, montonocity,
independence etc.) fail to condone a distribution of commodities $x$
extreme as that discussed above. By dint of its failure to maximise
social welfare $W\left(x\right)$. But to obtain this egalitarian
sensitivity to the distribution of income, three properties of Social
Welfare Functions are introduced. Which prove fatal to the a-politicality
of the economist's policy advice, and introduce presuppositions which
must lay naked upon the political passions of the economist, so much
more indecently for their hazy concealment under the technicalistic
canopy of functional mathematics.

Firstly, it is so famous a result as to be called the ``third theorem
of welfare economics'' that any such function $W\left(\cdot\right)$
as has certain ``uncontroversially'' desirable technical properties\footnote{As \citet{Sen1970} lists them: unrestricted domain in $S$, the ``Pareto
property'' (condoning Pareto improvements), and independence of irrelevant
alternatives in preference between any two alternatives in $S$.} will impose upon the polity $N$ the preferences of a dictator $i\in N$
within it. Arrow's famed ``impossibility'' theorem \citep{Arrow1951,Sen1970,Geanakoplos2005,Reny2001,Man2013}.
The preference of one individual $i\in N$ will serve to determine
the preference indicated between by society between different states
by $W\left(x\right)$. In practice, the preferences of the economist,
who decides upon the form of $W\left(\cdot\right)$ and thus imposes
their particular political passions (be they egalitarian or otherwise)
upon policy, deeming what is ``socially optimal'' by the different
weightings assigned to individual utilities $u_{i}\left(\cdot\right)$
within the polity. An excellent example of this is \citet{Diamond2011},
who demonstrate the ``social optimality'' of a 90\% top taxation
rate by assuming outright that the wellbeing of the wealthiest contributes
\emph{nothing} at the margin to social welfare\footnote{It is interesting to note that total confiscation is only not condoned
there as ``socially optimal'' for it does not trade losses from
avoidance behaviour against revenues raised as tax rates are increased
so as to maximise revenue from this source.}.

But the political presuppositions imported by the economist go deeper
in fact than this. Utilitarianism which allows for inter-personal
comparisons of utility in the construction of $W\left(x\right)$ requires
utility functions be ``cardinal'' - representing ``how much''
utility one derives from commodities over and above the bare preference
between different sets thereof. Utility is an \emph{extremely} vague
concept, because it was constructed to represent a common hedonistic
experiential metric where the very existence of such is uncertain
in the first place \citep{Fumagalli2013}. In practice, the economist
decides upon, extrapolates, assigns to $i\in N$ a particular utility
function which imports yet further assumptions about how any one individual
values their commodity allocation, and thus contributes to social
psychic wellbeing.

And finally, utilitarianism not only makes political statements about
who in the polity is to be assigned a disimproved situation. It makes
statements so outlandish and outrageous to the common sensibility
as to have provided the impetus for two of the great systems of philosophy
of justice in modernity - those of \citet{Rawls1971} and \citet{Sen1999,Sen2009}\footnote{The system of \citet{Dworkin1981,Dworkin1981a}, like that of \citet{Sen1999,Sen2009}
being developed as a constructive response to \citet{Rawls1971} too
could be construed as a reaction to the (to the common sensibility)
outlandish policies condoned by utilitarianism.}. Under almost any combination method $W\left(\cdot\right)$, the
maximisation of $W\left(\cdot\right)$ demands allocation to those
most able to realise utility from their allocation. It would demand,
for instance, redistribution of commodities from sick children to
the hedonistic libertine, for the latter can obtain greater ``utility''
there from. A problem so severe in its political implications it provided
the basic impetus for Rawls' and Sen's systems. \emph{A Theory of
Justice} is, of course, a direct response to the problematic political
content of utilitarianism.

So Pareto optimality stands as the best hope for the economist to
make a-political statements about policy, refraining from making statements
therein concerning the assignation of disimprovements in the situation
of any individual. Yet if applied to preferences over individual allocations
alone it condones some extreme situations of dubious political desirability
across the spectrum of political theory and philosophy. But how robust
a guide is it when we allow the polity to be concerned with states
of society in general? Not only their own individual allocation of
commodities. As they must be in the process of public reasoning in
every political philosophy from Plato to Popper and beyond. We will
see now, not at all. In all empirical situations Pareto optimality
offers no guide to policy-making, for policymaking must inevitably
make value judgements about who is to be assigned \emph{dis}improvement
in their situation.

\section{Under what conditions are Pareto improvements possible in economic
decision making about allocation?}

Let us now broaden our view to a weaker conception of Pareto one in
which we no longer restrict ourselves to assume that individuals care
only about their own allocation of goods and resources. Any economic
decision making is ultimately a decision to implement a movement between
two allocations $\left\{ x_{n}\right\} _{n\in N}\rightarrow\left\{ x_{n}'\right\} _{n\in N}$,
the question whether the associated movement between two states of
society $s\rightarrow s'$ associated with this movement between two
allocations is a Pareto improvement.

Because we are focussed on problems of economic decision making at
the societal level, let us suppose that the set of commodities $X\subset\mathbb{R}_{+}^{N_{x}}$
is contained within the set of information about society $X$ (so
that $X\subset S$), and that the allocation $\left\{ x_{n}\right\} _{n\in N}$
is contained within the set of information for any particular state
of $s$ (so that $\left\{ x_{n}\right\} _{n\in N}\subset s$). It
seems reasonable to suppose that $s$ will contain also any number
of transformations of this allocation $\left\{ x_{n}\right\} _{n\in N}$.
For instance, the statistical transformations which produce the summary
statistics of the allocation.

Let us now hold society outside of the economy constant so that we
may restrict ourselves to scenarios in which the preference-information
of individuals contains only some single-valued, individual-specific
transformation $f_{i}:\left\{ \left\{ x_{n}\right\} _{n\in N}\right\} \rightarrow\mathbb{R}$
of the possible allocations $\left\{ \left\{ x_{n}\right\} _{n\in N}\right\} \subset X$
of society. We might think of $f_{i}$ as representing something like
the process of reasoning applied by $i$ to the allocation $\left\{ x_{n}\right\} _{n\in N}$
of commodities in the economic system in order to arrive at that information
$f_{i}\left(\left\{ x_{n}\right\} _{n\in N}\right)$ on whose basis
they will form their preferences about the state of society. Let us
also suppose without great loss of generality that individual preferences
over that preference-information is monotonically increasing.

Hence for what follows we may effectively restrict our attention to
scenarios in which $\succeq_{i}$ is defined for $s_{i}=f_{i}\left(\left\{ x_{n}\right\} _{n\in N}\right)$
and monotonically increasing over the same.
\begin{defn}
If preferences are increasing over individual-specific transformations
of allocations of commodities $f_{i}$, and the non-economic state
of society is held constant (and thus effectively irrelevant), then
$f_{i}\left(\left\{ x_{n}\right\} _{n\in N}\right)\succeq f_{i}\left(\left\{ x_{n}'\right\} _{n\in N}\right)\iff f_{i}\left(\left\{ x_{n}\right\} _{n\in N}\right)\ge f_{i}\left(\left\{ x_{n}'\right\} _{n\in N}\right)$,
and $f_{i}\left(\left\{ x_{n}\right\} _{n\in N}\right)\succeq f_{i}\left(\left\{ x_{n}'\right\} _{n\in N}\right)\iff f_{i}\left(\left\{ x_{n}\right\} _{n\in N}\right)>f_{i}\left(\left\{ x_{n}'\right\} _{n\in N}\right)$.
\end{defn}
We can restate the definition of Pareto improvement for this class
of situations accordingly.
\begin{defn}
A movement between two states of society, $s\rightarrow s'$ is called
a \emph{Pareto improvement} if and only if $\exists i\in N:f_{i}\left(\left\{ x_{n}'\right\} _{n\in N}\right)\succ f_{i}\left(\left\{ x_{n}\right\} _{n\in N}\right)\,\&\,f_{j}\left(\left\{ x_{n}'\right\} _{n\in N}\right)\succeq f_{j}\left(\left\{ x_{n}\right\} _{n\in N}\right)\,\forall\,j\ne i\in N$.
\end{defn}
Neoclassical Pareto improvements are a special case of this definition,
specifically that special case where $x_{i}=f_{i}\left(\left\{ x_{n}\right\} _{n\in N}\right)\,\forall\,\left\{ \left\{ x_{n}\right\} _{n\in N}\right\} \subset X$.
While we are restricting our analysis here to changes in the economic
state of society, this restriction still models a relatively general
set of situations with respect to the individual preferences upon
which Pareto efficiency is predicated.

For instance, it is widely accepted in behavioural economics, and
has been for known for over a century \citep{Veblen1899,Duesenberry1949,Hirsch1977,Kahneman1979,Easterlin2001,Ariely2008,Clark2008,Frank2011,Layard2011,Barberis2013},
that individual preferences are not defined for absolute allocation
of commodities, but rather allocation relative to some reference point
or ``anchor''. Often, this reference point or anchor is other’s
consumption patterns, in which case we have, for instance

\[
f_{i}\left(\left\{ x_{n}\right\} _{n\in N}\right)=\frac{x_{i}}{x^{*}}
\]

The reference point $x^{*}$ may be the arithmetic mean of population
consumption, $\frac{1}{\left|N\right|}\sum_{n\in N}x_{n}$, or alternatively
the arithmetic mean over that portion of the population which is in
the ``neighbourhood'' of the individual in question\footnote{Technically speaking of course, if $x\in\mathbb{R}_{+}^{N_{x}}:N_{x}>1$
we really ought write this in linear algebraic form: $f_{i}\left(\left\{ x_{n}\right\} _{n\in N}\right)=x_{i}\left[x^{*}\right]^{-1}$.
The reference point expression would remain unchanged if it is the
arithmetic mean.}.

We may now establish when a movement between two states of society
constitutes a Pareto improvement in this context.
\begin{thm}
\label{prop:Possibility_of_Pimprovements}Suppose that we have a movement
between two states of society $\left\{ x_{n}\right\} _{n\in N}\rightarrow\left\{ x_{n}'\right\} _{n\in N}$
such that $\exists\left\{ i\right\} \subset N:x_{i}'>x_{i}$ and $x_{j}'\leq x_{j}\,\forall\,j\in N\setminus\left\{ i\right\} $,
and that individuals have monotonic preferences $\left\{ \succeq_{\cdot}\right\} _{\cdot\in N}$
over the individual-specific preference-information $f_{\cdot}\left(\left\{ x_{n}'\right\} _{n\in N}\right)$.
The movement is a Pareto improvement if and only if

\[
\frac{f_{k}\left(\left\{ x_{n}'\right\} _{n\in N}\right)-f_{k}\left(\left\{ x_{n}\right\} _{n\in N}\right)}{x_{i}'-x_{i}}\ge0\,\forall\,k\in N,\,i\in\left\{ i\right\} \subset N
\]

and

\[
\frac{f_{k}\left(\left\{ x_{n}'\right\} _{n\in N}\right)-f_{k}\left(\left\{ x_{n}\right\} _{n\in N}\right)}{x_{j}'-x_{j}}\leq0\,\forall\,k\in N,\,j\in N\setminus\left\{ i\right\} 
\]

with strict inequality in either case for at least one $k'\in N$.
\end{thm}
The conditions as sufficient are somewhat less interesting than they
are as necessary. If the conditions are \emph{not} met, the movement
between two states of society is not a Pareto improvement. If they
are \emph{not }met for every possible movement between two states
of society, then \emph{every} state of the world is a Pareto optimal
state.
\begin{cor}
\label{All_states_are_Pareto_optimal} If within the confines of the
conditions to which Theorem \ref{prop:Possibility_of_Pimprovements}
applies, the necessary and sufficient conditions for Pareto improvement
fail to hold for every movement $\left\{ x_{n}\right\} _{n\in N}\rightarrow\left\{ x_{n}'\right\} _{n\in N}$
between two states of society, then \emph{every }state of society
$\left\{ x_{n}\right\} _{n\in N}$ is Pareto optimal.
\end{cor}
Now let us consider what the necessary and sufficient conditions of
theorem \ref{prop:Possibility_of_Pimprovements} demand of each individual's
process of reasoning $f_{k}\left(\cdot\right)$ about the economic
state of society $\left\{ x_{n}\right\} _{n\in N}$. When $k\in\left\{ i\right\} \subset N$
and the individual is within the set of those who face an increased
allocation of commodities, they are fairly obvious, fairly reasonable
conditions. The first inequation states simply that the individual
$k$ form an assessment $f_{k}\left(\left\{ x_{n}'\right\} _{n\in N}\right)$
of the economic state of society $\left\{ x_{n}'\right\} _{n\in N}$
which is increasingly preferable or indifferent ``in'' (with respect
to, as a result of) the increase of their own increased allocation
of commodities, \emph{and} those of their peers within the set $\left\{ i\right\} \subset N$
of those who face an increased allocation of commodities\footnote{Though, as we will discuss below, this latter requirement even is
somewhat dubious}. And the second inequation requires that the individual $k$ form
an assessment $f_{k}\left(\left\{ x_{n}'\right\} _{n\in N}\right)$
of the economic state of society $\left\{ x_{n}'\right\} _{n\in N}$
which is indifferent or decreasingly preferable ``in'' (with respect
to, as a result of) the increase of the increased allocation of commodities
to those within the set $k\in N\setminus\left\{ i\right\} $ of those
who face decreased allocation of commodities. Which is essentially
(and rather crudely) to say that they must find the increase of their
own commodities desirable, and find indifferent or preferable the
increase or decrease of commodities to others as necessity for Pareto
improvement has it.

On the other hand, when we consider $k\in N\setminus\left\{ i\right\} $
and the individual is within the set of those who face an decreased
allocation of commodities, these conditions become both far more interesting,
and also rather far-fetched. The first inequation requires that the
individual $k$ form assessment $f_{k}\left(\left\{ x_{n}'\right\} _{n\in N}\right)$
of the economic state of society $\left\{ x_{n}'\right\} _{n\in N}$
which is increasingly preferable or indifferent ``in'' (with respect
to, as a result of) the increase of allocation to those of the individuals
within the set $\left\{ i\right\} \subset N$ of those who face an
increased allocation of commodities. The second ineqaution requires
that the individual $k$ form an assessment $f_{k}\left(\left\{ x_{n}'\right\} _{n\in N}\right)$
of the economic state of society $\left\{ x_{n}'\right\} _{n\in N}$
which is increasingly preferable or indifferent ``in'' (with respect
to, as a result of) the increase of their own \emph{decreased }allocation
of commodities, \emph{and} those of their peers within the set $N\setminus\left\{ i\right\} $
of those who face an weakly decreased allocation of commodities.

Now we see that these conditions are quite strong, to the extent that
we might humorously refer to them as the ``Kumbaya'', the ``hakuna
matata'' or ``blissful ignorance'' conditions. A polity characterised
by these conditions would be a utopian society. Literally. In the
sense that utopia stems from the ancient Greek for ``no-place''.
Or, more seriously, we might call them the ``universal, unconditional
altruism/ignorance'', or in a more sinister nomenclature the ``Brave
New World/concealment'' condition. They require, essentially, that
\emph{every} individual in society find it preferable, at least indifferent
to see some other individual acquire a increased commodity allocation
- become ``better off'' - if that is what is happening to that other
individual. Hence the necessity of ``universal, unconditional altruism'',
or ``ignorance''. But they also require at the same time that \emph{every}
individual in society find it preferable, or at least indifferent
that \emph{they themselves} or some other individual acquire a \emph{decreased}
commodity allocation - become ``worse off'' - if that is what is
happening to themselves or that other individual. Hence ``Brave New
World''\footnote{Recall the exquisitely disturbing conditioning spoken to genetically
engineered children grown in a test-tube as they sleep in \emph{Brave
New World}:
\begin{quote}
``Alpha children wear grey. They work much harder than we do, because
they're so frightfully clever. I'm really awfully glad I'm Beta, because
I don't work so hard. And then we are much better than the Gammas
and the Deltas. Gammas are stupid. They all wear green, and Delta
children wear khaki. Oh no, I \emph{don't} want to play with Delta
children. And Epsilons are too stupid to be able to read or write.
Besides, they wear black, which is such a beastly colour. I'm \emph{so
}glad I'm a Beta''.

\begin{flushright}
Aldous Huxley, \emph{Brave New World}, pp.24-25 (Flamingo Huxley Centenary
edition)
\par\end{flushright}\end{quote}
}, or ``concealment'', if the decreased allocation is to be concealed
from the necessary individuals to enforce by default their indifference.
Such a polity is at once the most ``Christian'' and the least ``Christian''
of nations (in the naive old fashioned sense of that word), for as
the necessity for Pareto optimality requires it, the movement of society
either inspires charitable feelings, pleasure at the dispossession
of others, or ignorance.

These conditions would outlaw the holding to by any in the polity
of the whole of Leftist politics \citep{Judt2010}, which most definitely
calls for not for a universal altruism, rather either an altruism
of the ``rich'' toward the ``poor'', or the coercion of the ``rich''
by the ``poor'' on the basis that the ``poor'' do \emph{not} find
increased commodity allocation to one group preferable. One is reminded
of the final few lines of \citet{Marx1848}:
\begin{quote}
``Let the ruling classes tremble at a Communistic revolution. The
proleterians have nothing to lose but their chains. They have a world
to win. WORKING MEN OF ALL COUNTRIES, UNITE!''
\end{quote}
They would also outlaw the holding to by any in the polity of the
whole of Rightist as well as the stronger liberal politics \citep{Mill1859,Strauss1953,Lucas1965,Nozick1974},
which would resist the wholesale coercion of the ``rich'' (or otherwise
``deserving'') in a redistribution of commodities away from them
toward the ``poor''. And most certainly anarchism, which would reject
as at all desirable \emph{any} coercion in the allocation of resources
\citep{Marshall1992}.

It is an empirical fact, already discussed, that assessments of the
economic state of society take a form similar to 
\[
f_{k}\left(\left\{ x_{n}\right\} _{n\in N}\right)=\frac{x_{k}}{x^{*}}
\]

indicating relativity of individual assessments of society to some
reference point. The reference point $x^{*}$ being, for instance,
the arithmetic mean of population consumption, $\frac{1}{\left|N\right|}\sum_{n\in N}x_{n}$.
Such that\footnote{Because if $f_{k}\left(\left\{ x_{n}\right\} _{n\in N}\right)=\frac{x_{k}}{x^{*}}$,
then $\frac{f_{k}\left(\left\{ x_{n}'\right\} _{n\in N}\right)-f_{k}\left(\left\{ x_{n}\right\} _{n\in N}\right)}{x^{*}-x^{*'}}\leq0$,
and if $x^{*}=\frac{1}{\left|N\right|}\sum_{n\in N}x_{n}$ then $\frac{x^{*}-x^{*'}}{x_{i}'-x_{i}}\geq0$
and so
\[
\frac{f_{k}\left(\left\{ x_{n}'\right\} _{n\in N}\right)-f_{k}\left(\left\{ x_{n}\right\} _{n\in N}\right)}{x^{*}-x^{*'}}\times\frac{x^{*}-x^{*'}}{x_{i}'-x_{i}}
\]
} if there is some movement in which $\exists i\in N:x_{i}'-x_{i}>0$
\[
\frac{f_{k}\left(\left\{ x_{n}'\right\} _{n\in N}\right)-f_{k}\left(\left\{ x_{n}\right\} _{n\in N}\right)}{x_{i}'-x_{i}}<0
\]

As others in society are allocated more by way of commodities, the
reference point rises, the relative standing of the individual deteriorates,
their assessment of society constitutes a \emph{dis}improvement.

It is quite easy to rationalise this empirical fact. It is well known,
and has been well known since \citet{Hirsch1977} that economic outcomes
depend on \emph{relative} standing in the distribution of acquired
commodities. The obtention of a job, the ability to obtain certain
commodities such as education at an elite school, indeed the obtention
of any commodity which is finite, all depend on the ability of the
individual to ``outbid'' others, and this in turn depends on their
relative standing in the distribution of commodities acquired. The
more others gain in their allocation, the more the individual's position
in the distribution deteriorates, and with it, their ability to obtain
commodities.

It is also a well known characteristic that the acquisition of commodities
reflects the selection within the evolutionary process in economies
of an increasingly (in the absence of any intervention or response
by competitors) dominant entity \citep{Nelson1982,Markey-Towler2014d},
whose economic dominance of other entities under certain conditions
only increases the more they are selected \citep{Markey-Towler2014d}.
And it is not mere conspiracy theory, but fact that concentration
of commodities to certain entities in the polity endows them with
political power as well as economic predominance \citep{Cardinale2014,Cardinale2015}.
As the evolutionary process increasingly allocates commodities to
an increasingly dominant entity, the ability of this entity to dominate
the polity through politics and economics increases at the expense
of the individual.

We can fairly safely conclude therefore that in empirical reality,
there is no movement between economic states of society which constitutes
a Pareto improvement. Unless all members of the polity are indifferent
to the movement (highly unlikely), there will always be at least one
individual who arrives, through their process of reasoning, at an
assessment of the movement as yielding a less preferable state of
society. And thus, by corollary \ref{All_states_are_Pareto_optimal},
all states of the world in empirical reality are Pareto optimal.

If every state of society is Pareto optimal, no policy can be implemented
which does not either leave all in the polity indifferent, or at least
one facing a \emph{dis}improvement in their assessment of the state
of society. Policies which cause a movement between economic states
of society, if they are to change anything at all with respect to
preferability, will \emph{necessarily} dispossess some individual
of a preferable assessment of the state of society. Economic policy
must therefore always statements about the assignation of \emph{dis}improvements
to this individual or that. Even if we restrict what constitutes a
political statement to statements which augur the deprivation of some
individual, assign to them \emph{dis}improvements, the formulation
and implementation of economic policy cannot therefore avoid making
political statements.

\section{The proper place of economics, and why it matters}

Is the concept of Pareto optimality robust? Does it have any value
as a criterion in the ``real'' empirical world? Does it offer us
a criterion for policy which does not make political statements, and
allow for economics to be divorced from political theory, and even
assert its priority and primacy therein?

The present work has demonstrated logico-mathematically, incontrovertibly,
that the answer is No. We are compelled inescapably by the mathematics
of Pareto optimality itself to recognise that in all empirical situations
economics cannot not even make political statements of a restricted
nature - about the assignation of\emph{ }``losses'' - let alone
of an unrestricted nature - making value judgements about the assignation
of ``gains''. This conclusion we arrived at by recognising that
when we allow the polity to form their assessments of the desirability
of social states, the empirical reality of those assessments means
that \emph{all} states of the world are Pareto optimal. There is no
policy to be implemented which affects a non-neutral change in the
economic states of society which does not assign \emph{dis}improvement
to some individual's assessment of the economic state of society.
Economics cannot be divorced from politics, and it the absolute primacy
of political theory and philosophy in the development and implementation
of policy must be recognised. We cannot escape the compulsion to embrace
political theory and philosophy as prior to any analysis of economic
policy.

So what? Why should we care? We should care because separating what
is economic science from what is political economy, firming as far
as possible the fuzzy boundary between fact and value \citep{Strauss1953},
seeking thereby ``objectivity'' is essential to a healthy political
sphere \citep{Sen1993}. The process of public reasoning is predicated
on there being some degree of objectivity in the views put forth therein
\citet{Sen2009}.

The democracies of the world sorely need a basic restoration of health
to their processes thereof, being in (and having been for some time
- see \citet{Habermas1962}) a crisis of superficial and corrosive
public discourse constituted by competing demagoguery on the part
of human mouthpieces for powerful and moneyed elements of the polity.
Demagoguery which will use whatever tools it can in desperation to
occupy a privileged place in the prejudices of the public, including
a false scientistic, \emph{faux} objective authority such as offered
by economic policy analysis proceeding on the basis of ``a-political'',
scientific economics guided by the search for Pareto optimality.

The fact that such authority is assumed by statements which are yet
political of necessity while appearing ostensibly not so is corrosive
to the public debate by obscuring what is fact and what is value,
and thereby usurping the authority which is due to political theory
and philosophy in the public debate. We have shown there are no ``ought''
statements to be derived by the economist devoid of political presuppositions.
Yet undergraduate economists are still taught the concept of Pareto
optimality as the basis for economic policy, professional economists
still utilise it in research, it still forms the basis for the ``proof''
that laissez-faire markets (corrected for ``imperfections'') are
``efficient'' or ``optimal''.

Still yet the argument we have made ought not be seen as purely negative.
It is as much an affirmation of the collaboration of Professors Sen
and Nussbaum, placing political theory and philosophy at the foundation
of welfare economics and thus obtaining the intellectual richness
contained within for economics, as it is a critique of economics.
It ought be read as encouragement for \emph{both} economists and political
scientists and philosophers.

Far better for the sake of the process of public reasoning that economists
recognise the absolute primacy and priority of political theory and
philosophy in the formulation and implementation of policy. As was
stated at the outset of this work, to continue to pretend otherwise
lends to the pronouncements of the economist a false scientistic authority
detrimental, even dangerous, for the process of public reasoning.
Far better for economists to engage fully with political theory and
philosophy in the manner of \citet{Sen1999,Sen2009} in developing
a new welfare economics. Expanding on the efforts of Professors Sen
and Nussbaum in particular to integrate into a system a set of an
intellectually rich, reasoned positions regarding the political theory,
and political philosophy of economics.

\section{Appendix: Proofs of Theorems}

\subsection{Proof of Theorem \ref{prop:Pareto_efficiency=00003DNo_Pareto_Improvements}}
\begin{proof}
(Necessity): Suppose, by way of contradiction, that there exists another
state $s'$ for which the movement $s\rightarrow s'$ is a Pareto
improvement. Then, by definition $\exists i\in N:s_{i}'\succ s_{i}\,\&\,s_{j}'\succeq s_{j}\,\forall\,j\ne i\in N$,
and so $\exists s'\in2^{S}\,\&\,i\in N:s_{i}'\succ s_{i}\,\&\,s_{j}'\succeq s_{j}\,\forall\,j\ne i\in N$.
But then $s$ could not be Pareto efficient. Hence a state $s\in2^{S}$
is Pareto efficient only if there is no other state $s'$ for which
the movement $s\rightarrow s'$ is a Pareto improvement.

(Sufficiency): If we can find no state $s'$ for which the movement
$s\rightarrow s'$ is a Pareto improvement, there exists no state
$s'$ for which $\exists i\in N:s_{i}'\succ s_{i}\,\&\,s_{j}'\succeq s_{j}\,\forall\,j\ne i\in N$.
Therefore $\nexists s'\in2^{S}\,\&\,i\in N:s_{i}'\succ s_{i}\,\&\,s_{j}'\succeq s_{j}\,\forall\,j\ne i\in N$,
and so $s$ is a Pareto efficient state. Hence a state $s$ is Pareto
efficient if there is no state $s'$ for which the movement $s\rightarrow s'$
would be a Pareto improvement.
\end{proof}

\subsection{Proof of Theorem \ref{Prop:The-extreme_case}}
\begin{proof}
The movement between two allocations, $\left\{ x_{i}\right\} _{i\in N}\rightarrow\left\{ x_{i}'\right\} _{i\in N}$
is a \emph{neoclassical Pareto improvement} if and only if $\exists i\in N:x_{i}'\succ x_{i}\,\&\,x_{j}'\succeq x_{j}\,\forall\,j\ne i\in N$.
Let us allocate more commodities to $j$ in a movement $\left\{ x_{i}\right\} _{i\in N}\rightarrow\left\{ x_{i}'\right\} _{i\in N}$
such that $x_{j}'>x_{j}$ while holding all other allocations constant,
so that $x_{i}=x_{i}'\,\forall\,i\ne j\in N$. Since individuals have
monotonically increasing preferences over only their own allocation,
$x_{j}'\succ_{j}x_{j}$, while $x_{i}\succeq_{i}x_{i}\,\forall\,i\ne j\in N$.
Hence the movement in question is a neoclassical Pareto improvement.
We can repeat the argument again to verify that another such movement
between allocations is a neoclassical Pareto improvement. This can
continue \emph{ad infinitum}, and the first argument is established.

Now suppose that the first allocation was neoclassical Pareto efficient.
If we now discover new commodities and allocate them entirely to individual
$j$, by the argument above we implement a neoclassical Pareto improvement.
But if we have now allocated the new commodities entirely to individual
$j$, the only movement between allocations in the absence of any
discovery of new commodities can be to redistribute the existing allocation.
Any such redistribution will entail a movement $\left\{ x_{i}\right\} _{i\in N}\rightarrow\left\{ x_{i}'\right\} _{i\in N}$
between allocations whereby $x_{j}'>x_{j}$ for at least one $j$
and $x_{i}'<x_{i}$ for at least one $i$. Since preferences are monotonically
increasing this means that $x_{j}'\succ_{j}x_{j}$ and $x_{i}'\prec x_{i}$,
hence $x_{i}'\npreceq x_{i}$, in which case this movement is not
a neoclassical Pareto improvement. Since this applies to any redistribution
of the existing allocation, no movement is a neoclassical Pareto improvement,
and by Theorem \ref{prop:Pareto_efficiency=00003DNo_Pareto_Improvements}
the allocation arrived at by allocating all newly discovered commodities
to $j$ is neoclassical Pareto efficient prior to any further discovery.
This establishes the second argument.
\end{proof}

\subsection{Proof of Theorem \ref{prop:Possibility_of_Pimprovements}}
\begin{proof}
The movement $\left\{ x_{n}\right\} _{n\in N}\rightarrow\left\{ x_{n}'\right\} _{n\in N}$
such that $\exists\left\{ i\right\} \subset N:x_{i}'>x_{i}$ and $x_{j}'\leq x_{j}\,\forall\,j\in N\setminus\left\{ i\right\} $
is a Pareto improvement if and only if
\[
f_{k}\left(\left\{ x_{n}'\right\} _{n\in N}\right)\succeq f_{k}\left(\left\{ x_{n}\right\} _{n\in N}\right)\,\forall\,k\in N
\]
\[
\&\,\exists k'\in N:f_{k'}\left(\left\{ x_{n}'\right\} _{n\in N}\right)\succ f_{k'}\left(\left\{ x_{n}\right\} _{n\in N}\right)
\]

Now if preferences are monotonically increasing over individual-specific
preference-information then we can say in fact that the movement will
be Pareto optimal if and only if 
\[
f_{k}\left(\left\{ x_{n}'\right\} _{n\in N}\right)-f_{k}\left(\left\{ x_{n}\right\} _{n\in N}\right)\ge0\,\forall\,k\in N
\]
\[
\&\,\exists k'\in N:f_{k'}\left(\left\{ x_{n}'\right\} _{n\in N}\right)-f_{k'}\left(\left\{ x_{n}\right\} _{n\in N}\right)>0
\]

This in hand we can demonstrate the necessary and sufficient conditions:

\emph{(Necessity)}: Suppose, by way of contradiction that

\[
\exists k\in n,\,i\in\left\{ i\right\} \subset N:\frac{f_{k}\left(\left\{ x_{n}'\right\} _{n\in N}\right)-f_{k}\left(\left\{ x_{n}\right\} _{n\in N}\right)}{x_{i}'-x_{i}}<0
\]

or

\[
\exists k\in n,\,j\in N\setminus\left\{ i\right\} :\frac{f_{k}\left(\left\{ x_{n}'\right\} _{n\in N}\right)-f_{k}\left(\left\{ x_{n}\right\} _{n\in N}\right)}{x_{j}'-x_{j}}>0
\]

or there is no strict inequality in either case for at least one $k'\in N$.
Take each case in turn. First, if 
\[
\exists k\in n,\,i\in\left\{ i\right\} \subset N:\frac{f_{k}\left(\left\{ x_{n}'\right\} _{n\in N}\right)-f_{k}\left(\left\{ x_{n}\right\} _{n\in N}\right)}{x_{i}'-x_{i}}<0
\]

Then as $\left\{ i\right\} \subset N:x_{i}'>x_{i}\implies x_{i}'-x_{i}>0$
we must have that $f_{k}\left(\left\{ x_{n}'\right\} _{n\in N}\right)-f_{k}\left(\left\{ x_{n}\right\} _{n\in N}\right)<0$,
which contradicts the movement $\left\{ x_{n}\right\} _{n\in N}\rightarrow\left\{ x_{n}'\right\} _{n\in N}$
being a Pareto improvement. Second, if
\[
\exists k\in n,\,j\in N\setminus\left\{ i\right\} :\frac{f_{k}\left(\left\{ x_{n}'\right\} _{n\in N}\right)-f_{k}\left(\left\{ x_{n}\right\} _{n\in N}\right)}{x_{j}'-x_{j}}>0
\]

Then as $x_{j}'\leq x_{j}\,\forall\,j\in N\setminus\left\{ i\right\} \implies x_{j}'-x_{j}\leq0$
we have must that $f_{k}\left(\left\{ x_{n}'\right\} _{n\in N}\right)-f_{k}\left(\left\{ x_{n}\right\} _{n\in N}\right)<0$,
which contradicts the movement $\left\{ x_{n}\right\} _{n\in N}\rightarrow\left\{ x_{n}'\right\} _{n\in N}$
being a Pareto improvement. Finally, suppose there is no strict inequality
in either case for at least one $k'\in N$, so that
\[
\frac{f_{k}\left(\left\{ x_{n}'\right\} _{n\in N}\right)-f_{k}\left(\left\{ x_{n}\right\} _{n\in N}\right)}{x_{i}'-x_{i}}=0\,\forall\,k\in N,\,i\in\left\{ i\right\} \subset N
\]

and

\[
\frac{f_{k}\left(\left\{ x_{n}'\right\} _{n\in N}\right)-f_{k}\left(\left\{ x_{n}\right\} _{n\in N}\right)}{x_{j}'-x_{j}}=0\,\forall\,k\in N,\,j\in N\setminus\left\{ i\right\} 
\]

Or, collapsing these to one expression:
\[
\frac{f_{k}\left(\left\{ x_{n}'\right\} _{n\in N}\right)-f_{k}\left(\left\{ x_{n}\right\} _{n\in N}\right)}{x_{n}'-x_{n}}=0\,\forall\,k,n\in N
\]

But then $f_{k}\left(\left\{ x_{n}'\right\} _{n\in N}\right)-f_{k}\left(\left\{ x_{n}\right\} _{n\in N}\right)=0\,\forall\,k,n\in N$,
which contradicts the necessity of there being at least one $k'\in N:f_{k'}\left(\left\{ x_{n}'\right\} _{n\in N}\right)-f_{k'}\left(\left\{ x_{n}\right\} _{n\in N}\right)>0$
for the movement $\left\{ x_{n}\right\} _{n\in N}\rightarrow\left\{ x_{n}'\right\} _{n\in N}$
to be a Pareto improvement .

\emph{(Sufficiency)}: Suppose we have
\[
\frac{f_{k}\left(\left\{ x_{n}'\right\} _{n\in N}\right)-f_{k}\left(\left\{ x_{n}\right\} _{n\in N}\right)}{x_{i}'-x_{i}}\ge0\,\forall\,k\in N,\,i\in\left\{ i\right\} \subset N
\]

and

\[
\frac{f_{k}\left(\left\{ x_{n}'\right\} _{n\in N}\right)-f_{k}\left(\left\{ x_{n}\right\} _{n\in N}\right)}{x_{j}'-x_{j}}\leq0\,\forall\,k\in N,\,j\in N\setminus\left\{ i\right\} 
\]

with strict inequality in either case for at least one $k'\in N$.
Then as $\left\{ i\right\} \subset N:x_{i}'>x_{i}\implies x_{i}'-x_{i}>0$
we have that
\[
f_{k}\left(\left\{ x_{n}'\right\} _{n\in N}\right)-f_{k}\left(\left\{ x_{n}\right\} _{n\in N}\right)\ge0\,\forall\,k\in N,\,i\in\left\{ i\right\} \subset N
\]

And as we have $x_{j}'\leq x_{j}\,\forall\,j\in N\setminus\left\{ i\right\} \implies x_{j}'-x_{j}\leq0$
we have that
\[
f_{k}\left(\left\{ x_{n}'\right\} _{n\in N}\right)-f_{k}\left(\left\{ x_{n}\right\} _{n\in N}\right)\geq0\,\forall\,k\in N,\,j\in N\setminus\left\{ i\right\} 
\]

and with strict inequality in either case for at least one $k'\in N$.
Which confirms the sufficient conditions for the movement $\left\{ x_{n}\right\} _{n\in N}\rightarrow\left\{ x_{n}'\right\} _{n\in N}$
to be a Pareto improvement.
\end{proof}

\subsection{Proof of corollary \ref{All_states_are_Pareto_optimal}}
\begin{proof}
If the conditions to which Theorem \ref{prop:Possibility_of_Pimprovements}
are the case, and the necessary and sufficient conditions identified
by that theorem for Pareto improvement fail to hold then by that theorem,
because they are necessary, there is no Pareto improvement in that
movement. If those conditions fail to hold for every movement between
two states of the world $\left\{ x_{n}\right\} _{n\in N}\rightarrow\left\{ x_{n}'\right\} _{n\in N}$
, then there is no Pareto improvement to be made by movement from
any and every state of the world $\left\{ x_{n}\right\} _{n\in N}$.
Thus by theorem \ref{prop:Pareto_efficiency=00003DNo_Pareto_Improvements},
this is sufficient (and necessary) for every state of the world $\left\{ x_{n}\right\} _{n\in N}$
to be Pareto optimal.
\end{proof}
\bibliographystyle{elsarticle-harv}
\bibliography{ReferenceDatabase}

\begin{thebibliography}{37}
\expandafter\ifx\csname natexlab\endcsname\relax\def\natexlab#1{#1}\fi
\expandafter\ifx\csname url\endcsname\relax
  \def\url#1{\texttt{#1}}\fi
\expandafter\ifx\csname urlprefix\endcsname\relax\def\urlprefix{URL }\fi

\bibitem[{Ariely(2008)}]{Ariely2008}
Ariely, D., 2008. Predictably Irrational. Harper Perennial, New York.

\bibitem[{Arrow(1951)}]{Arrow1951}
Arrow, K., 1951. Social choice and individual values. Wiley, New York.

\bibitem[{Barberis(2013)}]{Barberis2013}
Barberis, N., 2013. Thirty years of prospect theory in economics. Journal of
  Economic Perspectives 27~(1), 173--196.

\bibitem[{Cardinale(2015)}]{Cardinale2015}
Cardinale, I., 2015. Resources, Production and Structural Dynamics. Cambridge
  University Press, Cambridged, Ch. Towards a political economy of resources,
  pp. 198--210.

\bibitem[{Cardinale and Coffman(2014)}]{Cardinale2014}
Cardinale, I., Coffman, D., 2014. Economic interdependencies and political
  conflict: The political economy of taxation in eighteenth century {Britain}.
  Economia Politica XXXI~(3), 277--300.

\bibitem[{Clark et~al.(2008)Clark, Frijters, and Shields}]{Clark2008}
Clark, A., Frijters, P., Shields, M., 2008. Relative income, happiness and
  utility: An explanation for the {Easterlin} paradox and other puzzles.
  Journal of Economic Literature 46~(1), 95--144.

\bibitem[{Diamond and Saez(2011)}]{Diamond2011}
Diamond, P., Saez, E., 2011. The case for a progressive tax: From basic
  research to policy recommendations. Journal of Economic Perspectives 25~(4),
  165--190.

\bibitem[{Duesenberry(1949)}]{Duesenberry1949}
Duesenberry, J., 1949. Income, Saving, and the Theory of Consumer Behavior.
  Harvard University Press, Cambridge, Massachusetts.

\bibitem[{Dworkin(1981{\natexlab{a}})}]{Dworkin1981}
Dworkin, R., 1981{\natexlab{a}}. What is equality? part 1: Equality of welfare.
  Philosophy and Public Affairs 10~(3), 185--246.

\bibitem[{Dworkin(1981{\natexlab{b}})}]{Dworkin1981a}
Dworkin, R., 1981{\natexlab{b}}. What is equality? part 2: Equality of
  resources. Philosophy and Public Affairs 10~(4), 283--345.

\bibitem[{Easterlin(2001)}]{Easterlin2001}
Easterlin, R., 2001. Income and happiness: Toward a unified theory. Economic
  Journal 111, 465--484.

\bibitem[{Frank(2011)}]{Frank2011}
Frank, R., 2011. The {Darwin} Economy. Princeton University Press, Princeton.

\bibitem[{Fumagalli(2013)}]{Fumagalli2013}
Fumagalli, R., 2013. The futile search for true utility. Economics and
  Philosophy 29~(3), 325--347.

\bibitem[{Geanakoplos(2005)}]{Geanakoplos2005}
Geanakoplos, J., 2005. Three brief proofs of {Arrow's} impossibility theorem.
  Economic Theory 26~(1), 211--215.

\bibitem[{Habermas(1962)}]{Habermas1962}
Habermas, J., 1962. The Structural Transformation of the Public Sphere. Polity,
  Cambridge.

\bibitem[{Hirsch(1977)}]{Hirsch1977}
Hirsch, F., 1977. The Social Limits to Growth. Routledge, London.

\bibitem[{Judt(2010)}]{Judt2010}
Judt, T., 2010. Ill Fares the Land. Penguin, London.

\bibitem[{Kahneman and Tversky(1979)}]{Kahneman1979}
Kahneman, D., Tversky, A., 1979. Prospect theory: An analysis of decision under
  risk. Econometrica 47~(2), 263--292.

\bibitem[{Layard(2011)}]{Layard2011}
Layard, R., 2011. Happiness, revised Edition. Penguin, New York.

\bibitem[{Lucas(1965)}]{Lucas1965}
Lucas, J., 1965. Against equality. Philosophy 40~(154), 296--307.

\bibitem[{Man and Takayama(2013)}]{Man2013}
Man, P., Takayama, S., 2013. A unifying impossibility theorem. Economic Theory
  54~(2), 249--271.

\bibitem[{Marshall(1992)}]{Marshall1992}
Marshall, P., 1992. Demanding the Impossible, A History of Anarchism. Harper
  Perennial, London.

\bibitem[{Marx and Engels(1848)}]{Marx1848}
Marx, K., Engels, F., 1848. The Communist Manifesto. Penguin, London.

\bibitem[{Mas-Collel et~al.(1995)Mas-Collel, Winston, and
  Green}]{Mas-Collel1995}
Mas-Collel, A., Winston, M., Green, J., 1995. Microeconomic Theory. Oxford
  University Press, Oxford.

\bibitem[{McCloskey(1983)}]{McCloskey1983}
McCloskey, D., 1983. The rhetoric of economics. Journal of Economic Literature
  21~(2), 481--517.

\bibitem[{Mill(1859)}]{Mill1859}
Mill, J., 1859. On Liberty. Penguin, London.

\bibitem[{Nelson and Winter(1982)}]{Nelson1982}
Nelson, R., Winter, S., 1982. An Evolutionary Theory of Economic Change.
  Belknap Harvard University Press, Cambridge, Massachussetts.

\bibitem[{Nozick(1974)}]{Nozick1974}
Nozick, R., 1974. Anarchy, State and Utopia. Basic Books, New York.

\bibitem[{Rawls(1971)}]{Rawls1971}
Rawls, J., 1971. A Theory of Justice. Belknap Harvard University Press,
  Cambridge, Massachussetts.

\bibitem[{Reny(2001)}]{Reny2001}
Reny, P., 2001. {Arrow's} theorem and the {Gibbard-Satterthwaite} theorem: a
  unified approach. Economics Letters 70, 99--105.

\bibitem[{Sen(1970)}]{Sen1970}
Sen, A., 1970. Collective Choice and Social Welfare. North-Holland, Amsterdam.

\bibitem[{Sen(1973)}]{Sen1973}
Sen, A., 1973. On Economic Inequality. Oxford University Press, Oxford.

\bibitem[{Sen(1993)}]{Sen1993}
Sen, A., 1993. Positional objectivity. Philosophy and Public Affairs 22~(2),
  126--145.

\bibitem[{Sen(1999)}]{Sen1999}
Sen, A., 1999. Commodities and Capabilities. Oxford University Press, Oxford.

\bibitem[{Sen(2009)}]{Sen2009}
Sen, A., 2009. The Idea of Justice. Harvard University Press, Cambridge,
  Massachusetts.

\bibitem[{Strauss(1953)}]{Strauss1953}
Strauss, L., 1953. Natural Right and History. University of Chicago Press,
  Chicago.

\bibitem[{Veblen(1899)}]{Veblen1899}
Veblen, T., 1899. The Theory of the Leisure Class. Oxford University Press,
  Oxford World's Classics.

\end{thebibliography}

\end{document}